# Increasing THz Radiation Power and Monochromaticity Using Optically Induced Photonic Crystal in Layered Superconductors

Alireza Kokabi[1], Hamed Kamrani[1], Mehdi Fardmanesh[1]

**Abstract:** The possibility of enhancing the radiation power and monochromaticity by optically induced photonic crystal in the superconducting cavity is proposed and investigated. In such a structure, by periodically irradiating the stacked Josephson junctions and consequently partially suppression of the superconductivity in the irradiated positions due to depairing, a periodic optical configuration is formed. This leads to photonic band gap opening in the range of the terahertz radiation emitted from the layered superconductor. We show that such a photonic band gap significantly enhances the impedance matching at the boundary of the cavity and the waveguide. Since the weak optical coupling of the outer and inner space of layered superconductor samples is a serious reason of reducing radiated power especially in the experiments, the proposed configuration is capable of extremely enhancement in the emitted power along with attenuation of the undesired harmonics.

*Index Terms*—Terahertz oscillators, Stack of intrinsic Josephson junctions, Photonic band gap

## 1-Introduction

Continuous-wave terahertz sources are the topic of some recent tremendously high frequency (THF) device oriented works due to the technology demands for the utilization of the terahertz band in the materials science, biology, security checking, and so on [1]. High power emissions of 1mW are therefore desirable for realizing compact THz source applications.

The emission in the terahertz range for the $Bi_2Sr_2CaCu_2O_{8+\delta}$ (Bi-2212) is possible since the size of the superconducting gap is higher than the Josephson plasma edge. Such a radiation occurs in the presence of applied magnetic field by the proposed mechanism of the flux-flow radiation and in the absence of the magnetic field due to the ac Josephson effect [2-12]. The emitted frequency is proportional to the applied voltage and inversely proportional to the width of the rectangular mesa.

The concept of increasing radiation power is one of the most serious investigated fields since the range of the applications is extremely extended when the obtained power is enhanced. The monochromaticity and emission stability of such radiations are the other concerning issues, which enable the large number of single harmonic applications.

Nevertheless, the obtained radiation power is still a very low fraction of supplying electric power as for instance the emission power in the order of 0.5-5μW corresponding to the fraction of about 0.003-0.03% have been reported in recent years [3, 13]. Accordingly, there exists a lot of capability for increasing the amount of emitting electromagnetic power. To benefit this large capability, some modifications should be applied to improve the mentioned fraction and enhance the emission power due to either of ac Josephson and cavity resonance. One of these modifications might be entering *I-V* characteristics and phase dynamics into the nonlinear state. The optimal state for the radiation is the one with nonlinear *I-V* characteristics where a large part of the input power can be pumped into plasma oscillation. In this regard, working with larger junctions increases the nonlinearity; however, extended path inside the lossy cavity makes the dissipation more dominant.

Utilization of other possible substrate shapes and materials such as metallic ones, removing the substrate, and using a surrounding external cavity could be other desired modifications [14-16]. However, these approaches might introduce some difficulties in the fabrication and characterization of the mesa structures.

On the other hand, improving the optical coupling of the cavity to the outside propagation media by the impedance matching at the boundary could result in the significant enhancement of the radiation intensity. It has been pointed out that there exists a large impedance mismatch between the intrinsic Josephson junctions and the outside space due to the small ratio between the thickness of the stack and the penetration depth [8]. The effect of impedance mismatch on the fraction of radiated power with respect to the total biasing power can be estimated by the proposed models for the boundaries of strip lines and waveguides [17]:

$$P_{rad}/P_{tot} = \frac{4Z_0 Z_s}{(Z_0 + Z_s)^2} \approx \sqrt{\frac{8\lambda d}{\varepsilon w^2}} \qquad (1)$$

where $Z_0$ is the waveguide impedance, $Z_s$ is the strip line impedance in the limit $k_\omega w \ll 1$ and $Z_s \ll Z_0$, $\lambda$ is the London penetration length of the superconducting leads, $d$ is the thickness of the insulating layer, $\varepsilon$ is the dielectric constant, $k_\omega = \omega/c$ and $w$ is the junction width. According to this equation, the weak coupling leads to power efficiency of $10^{-5}$, which is consistent with the experimental data. This analysis also shows that the low fraction of the radiated power can mainly be attributed to the impedance mismatch at the radiation boundary of intrinsic Josephson junction. Successive reflections from the boundaries due to the imperfect coupling to the outer space, which prolongs the existence of the radiation inside the lossy cavity, might be an explanation for such dissipation. Despite this importance of optical coupling, no breakthrough progress has been reported in this regard.

In addition to the above analysis, there exist numerical investigations of the inherent impedance matching at the sample boundaries. As reported in Ref. [10] decreasing the Z from 10000 to 1000 may increase the radiation power. The surface impedance dependence of the radiation intensity is also discussed in [9] while the obtained results show that decreasing Z in the range of 10000 to 100 for first three cavity modes leads to enhanced intensity and the strongest emission is observed for Z in the order of 50 to 100.

Based on such results, it seems that the effect of surface impedance might be very serious since it can significantly improve the power efficiency. In addition, controlling this impedance can be an interesting way of tuning the emitted power. Nevertheless, the problem is to make change in the surface impedance without imposing remarkable effect on the emission behavior of the intrinsic junctions. Anyway, it seems that the concept of photonic crystals is a suitable choice that can be applied in this case for imposing change in the coupling of the stacked junctions at their boundary to the outside.

## 2-Theoretical Approach

Here, we propose inducing photonic band gap in the cavity of the intrinsic Josephson junctions using *c*-axis oriented patterned optical irradiation. While light patterning is a crucial part of the modern nanotechnology fabrication processes, it seems that it could also play a main role in the induction of the desired photonic band gap in the hosting superconducting media. In this approach, the superconductivity is partially suppressed through the process of the cooper-pair breaking caused by the photon absorption in some regions. In the other words, the terahertz emitting mesa structure simultaneously behaves as a photonic crystal by the incident periodic radiation pattern. The dispersive permittivity of the superconductors by the approach of the Drude model, obtained from the following equation [18-20]:

$$\varepsilon_{eff} = \varepsilon_c \left\{ 1 - \frac{\Omega_{sp}^2(x,y)}{\omega^2} - \frac{\Omega_{np}^2(x,y)}{\omega[\omega + i\gamma(x,y)]} \right\}, \qquad (2)$$

in which $\Omega_{sp}=(n_s e^2/m_e \varepsilon)^{1/2}$ and $\Omega_{np}=(n_n e^2/m_e \varepsilon)^{1/2}$ are the plasma frequencies corresponding to the superconducting and quasiparticle carriers, respectively. Applying striped periodic illumination leads to the spatial variation of the permittivity inside the superconductor and opens the photonic band gap. It is remarkable that the frequency of the patterning radiation should be higher than that of the superconductivity specific gap, $2\Delta$, in order that it could partially diminish the superconductivity. On the other hand, the amount of the superconductivity suppression, which is controlled by the illumination intensity, should not be destructive to the plasma oscillation of the intrinsic Josephson junctions.

Among the large number of models proposed for the photo-effect in the superconductors, here, we apply $T^*$ model [21] to estimate the number of quasiparticles generated by photon absorption, which seems more suitable in this case due to the better steady-state modeling. The relation between excess number density of quasiparticles, $\Delta n_n$, and absorbed optical energy can be expressed as [19]

$$\Delta n_n = n_n \left[ \left( 1 + \frac{F\tau_{eff}}{n_n E} P \right)^{1/2} - 1 \right], \qquad (3)$$

where $n_n$ is the thermal equilibrium number density of quasiparticles, $F$ is the fraction of the absorbed optical energy that is shared among the excess quasiparticles, $\tau_{eff}$ is the effective recombination time, $E$ is the average energy of an excited quasiparticle, and $P$ is the absorbed energy flux per unit volume.

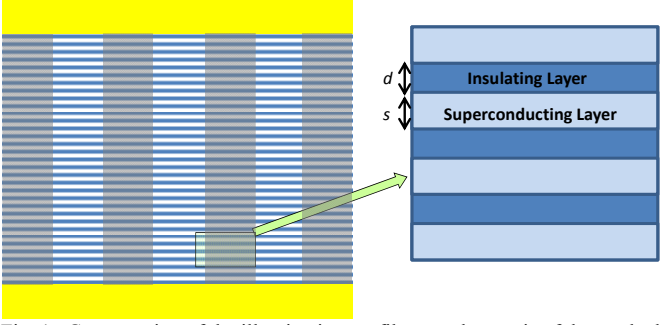

Fig. 1. Cross section of the illumination profile over the *c*-axis of the stacked junctions. The bright and dark strips show irradiated and non-irradiated regions.

Figure 1 shows the cross section of illumination pattern over the entire sample for a rectangular-shape mesa. As illustrated in this figure, the light pattern is a periodic configuration in which some regions are not illuminated and the remaining parts are partially irradiated. These two parts are clarified by vertical dark and bright strips in this figure. It should be mentioned that for other types of mesa structures such as cylindrical one, 2D periodic illumination pattern is suggested to impose photonic band gap in all in-plane directions. Here, we neglect the effect of intensity decay due to the light absorption in the *c*-axis of the mesa structure. This might be an acceptable approximation since the thickness of the practical mesa structures is in the order of micrometer and on the other hand a wide range of light wavelengths capable of depairing can be applicable here. Thus, one could select a special wavelength for irradiation with minimum absorption coefficient inside the Bi-2212 material.

Assuming the structure drawn in figure 1 is infinitely extended in both directions, its photonic band structure obtained based on the dispersive optical transfer-matrix method is presented in figure 2. The TE mode band structure shows a large cutoff frequency at the frequency of 2.2THz, which is appropriate for the first three cavity modes of the radiation. It is remarkable that the existence of the cutoff frequency for the periodic structure of Abrikosov lattice is also observed previously [19,20]. For obtaining such a cutoff frequency, the widths of the dark and illuminated regions are both set to be 2.65μm. With infinite successive dark and bright regions in one side of an assumptive boundary, the reflection of an outgoing electromagnetic wave incident to the waveguide boundary would be negligible. Although the cavity is bounded, striped lightening can affect on the reflection of the boundaries and consequently change the impedance of the stacked junctions. Since the periodicity of this light patterning is 5.3μm, about 18 periods can be embedded inside a cavity of 100μm. Using the modern illumination facilities, which are usually applied in nanotechnology fabrication methods such as photolithography, this light patterning configuration might not be serious obstacle in the proposed method especially with the required minimum feature size of micro-meter. Such a large minimum feature size allows for the utilization of a wide wavelength range with negligible diffraction issues. This number of periods is sufficient for suppressing a considerable amount of boundary reflections.

The impedance of a waveguide is the ratio of the transverse electric and magnetic fields,

$$Z = \frac{E_\perp}{H_\perp}. \qquad (4)$$

In the case of photonic crystals, the components of the oscillating fields for the TEM propagation are also needed. Among the investigated methods for the impedance of the photonic crystal, here the method presented in Ref. [22,23] is applied. The results of the impedance variations at different radiation intensities are presented in table 1, which is obtained numerically by the finite-element method (FEM). The data are obtained by the assumption of the unity permeability, which only scales the considered inherent impedances and does not affect on the results. The obtained results show that increasing light intensity would decrease the difference between the inherent impedance of the cavity and the vacuum waveguide and extensively enhance the matching to the outer space. Similar results are obtainable by the assumption of different waveguide impedances. Accordingly, the emission power and the boundary coupling can also be tuned by the light intensity.

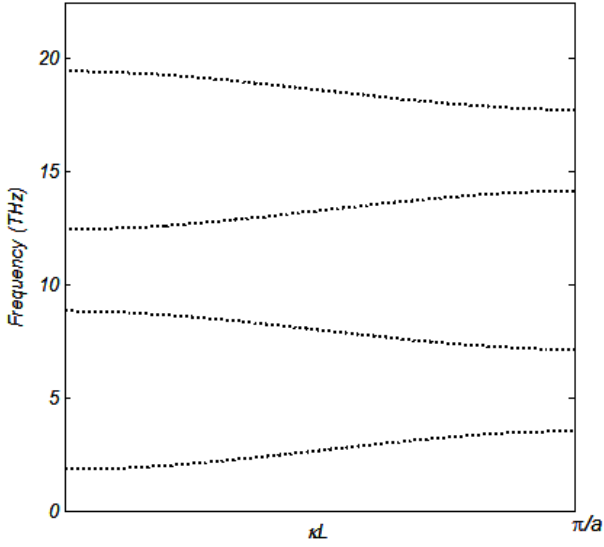

Fig. 2. TE band structure of the irradiated junctions in one dimension.

TABLE I
VARIATION OF CAVITY IMPEDANCE BY PATTERNED ILLUMINATION

| Percentage of $n_s$ suppressed by illumination | Cavity Impedance ($\Omega$) | Vacuum Impedance ($\Omega$) |
|---|---|---|
| 0% | 120 | 377 |
| 10% | 148 | 377 |
| 20% | 203 | 377 |
| 30% | 271 | 377 |
| 40% | 332 | 377 |

The monochromaticity of the emitted power is also of great importance. We apply the well-known sine-Gordon equations to investigate the effects of the patterned illumination on the power spectrum of the electromagnetic emission in the presence and absence of external magnetic field as follow [9]:

$$\frac{\partial^2 \varphi_{l+1,l}}{\partial \chi^2} = \left(1 - \zeta \Delta_n^2\right)\left(\frac{\partial \xi_{l+1,l}}{\partial \tau} + \beta \xi_{l+1,l} + \sin \varphi_{l+1,l} - j\right) \quad (5)$$

$$\frac{\partial \varphi_{l+1,l}}{\partial \tau} = \left(1 - \alpha \Delta_n^2\right) \xi_{l+1,l},$$

in which the finite difference operator is defined as $\Delta_n^2 f_n = f_{n+1} + f_{n-1} - 2f_n$ and operates on the gauge invariant phase difference, $\varphi_{l+1,l}$, and electric field, $\xi_{l+1,l}$, at the $l$-th junction, the parameters $\alpha = \varepsilon_c \mu^2 / \varepsilon_0 s d$ and $\zeta = \lambda_{ab}^2 / s d$ are capacitive and inductive couplings between different junctions respectively, and $\beta = \sigma_c \lambda_c / c(\varepsilon_0 \varepsilon_c)^{1/2}$ is the damping term. In these relations, $s$ is the thickness of the superconducting layers, $\sigma_c$ is c-axis conductivity, $\varepsilon_c$ is the permittivity of the junctions, and $\mu$ is the Debye screening length. The dynamic boundary condition [12] is applied for the coupled equations as follow.

$$\frac{\partial \varphi_{l+1,l}}{\partial \chi} = \langle b_{l+1,l} \rangle_\tau + \tilde{b}_{l+1,l}$$

$$\frac{\partial \varphi_{l+1,l}}{\partial \tau} = j/\beta + \tilde{\xi}_{l+1,l} \qquad (6)$$

$$\tilde{b}_{l+1,l} = \pm \sqrt{\varepsilon_c/\varepsilon_d}\, \tilde{\xi}_{l+1,l},$$

where $z=(\varepsilon_c/\varepsilon_d)^{1/2}$ models the impedance mismatch between the cavity and the outside dielectric, $\varepsilon_d$ is the relative permittivity of the dielectric, $b_{l+1,l}$ is the magnetic field, the symbol $\langle\ldots\rangle_\tau$ stands for time averaging, and the symbol ~ over field quantities represents their oscillatory parts. More details of numerical calculations and the effects of the external magnetic field can be obtained in Ref. [24].

The London penetration depths along the *c*-axis and CuO planes, $\lambda_c$ and $\lambda_{ab}$, and the *c*-axis conductivity, $\sigma_c$, depend on the paired carrier density through the following equations:

$$\lambda_c = \lambda_{c0}\sqrt{\frac{n_{s0}}{n_{s0}+\Delta n_s}},\ \sigma_c = \sigma_{c0}\frac{n_{n0}+\Delta n_n}{n_{n0}}. \qquad (7)$$

Here, we assume that the light illumination changes the values of $\lambda_c$ and $\lambda_{ab}$, and, $\sigma_c$ in the bright regions while it has no effect on the dark regions. The continuity condition between the dark and bright regions is applied due to the existence of phase difference spatial derivative in the equation and the tangential component continuity of electric field.

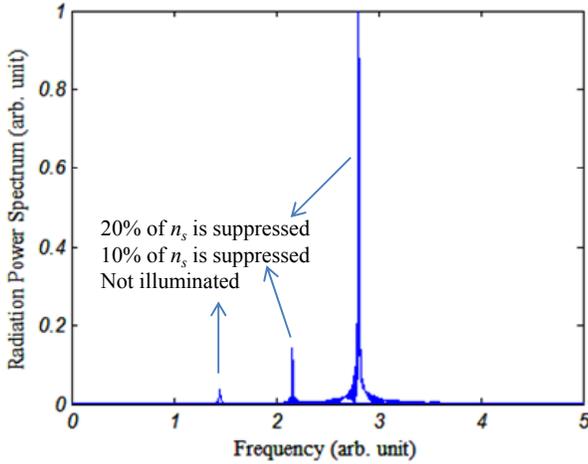

Fig. 3. Three different spectrums obtained by three amounts of illumination intensities.

## 3-Results and Discussions

The power spectrums for three different light intensities without external magnetic bias are plotted in figure 3. These results are obtained using the impedances presented in table 1 as the boundary conditions of equation 5. The intensities are expressed in terms of superconducting carrier suppression using equation 3. For better comparison and visibility, the spectrums are laterally shifted to avoid overlap between coinciding peaks. As observed in this figure, the sharpness of the emission is not considerably affected by the applied patterned illumination, which means that the impedance tuning becomes possible without undesired effects such as widening the power spectrum or emitting secondary harmonics. The results also show a large amplification in the emission power by applying striped lightening. For instance, assuming that the light intensity is capable of depairing 20% of $n_s$ in the illuminated regions, the radiation power is amplified more than one order of magnitude with respect to that of the non-irradiated mesa. It is remarkable that similar results are also obtainable by applying the constant magnetic field in equation 5.

The cutoff frequency can also provide the possibility of attenuating the higher radiating harmonics and consequently improve the monochromaticity of the output electromagnetic power. To this purpose, the cutoff frequency should be below the undesired harmonics in order to maintain them out of the induced photonic gap. This strategy leads to extensive enhancement of the surface impedance for the desired harmonics while this impedance remains almost unchanged or even gets worse for the undesired ones. Even in the case of the non-cavity mode radiation frequencies, which have been observed recently in some experimental works [14], such a strategy is still applicable for the attenuation of the undesired modes provided that the cutoff frequency would be adjusted below them.

The inherent impedance for the first and third harmonics at the presence and the absence of the patterned illumination are presented in table 2. The periodicity of the lightening pattern is set to be 8μm in order to achieve the cutoff frequency of 0.73THz. This means that only the first harmonic is inside of the photonic band gap. The obtained data confirm that the impedance variation for the first harmonic is significantly larger than that of the third harmonic.

TABLE II
VARIATION OF CAVITY IMPEDANCE BY PATTERNED ILLUMINATION

| Percentage of $n_s$ suppressed by illumination | Cavity Impedance for First Harmonic ($\Omega$) | Cavity Impedance for Third Harmonic ($\Omega$) | Vacuum Impedance ($\Omega$) |
|---|---|---|---|
| 0% | 120 | 115 | 377 |
| 10% | 151 | 134 | 377 |
| 20% | 182 | 121 | 377 |
| 30% | 286 | 112 | 377 |

Finally, it should be mentioned that the partially suppression of the superconductivity might increase the Joule heating effects and make some difficulties in the temperature control of the mesa structure in the experimental works. In this regard, increasing the thermal coupling between the substrate or mesa structure and the cold-head might compensate the excess of the heat flux due to the decreased *c*-axis conductivity and resolve the problem in many cases. Nevertheless, there would be the capability of forming hot-spots inside the cavity at the illuminated regions. This effect might result in the creation of the synchronized standing-waves between two neighboring hot-spots and generate new resonance modes of radiation, which can also enhance the emission power. Similar phenomenon is theoretically and experimentally observed at high bias currents without optical illuminations [25-27]. This possibility is under further investigations.

## 4- Conclusion

We applied the striped light illumination to enhance the optical coupling of the intrinsic Josephson junction stack to the outer space and consequently improve the radiation capability of such devices. The results of numerical analyses show that this proposed configuration increases the emission intensity while preserves the considered original radiating properties of the stacked junctions. It was also shown that the illumination intensity and periodicity could tune the surface impedance and consequently the emission power. The proposed method can also attenuate the higher harmonics and increase the radiation monochromaticity.